# Direct Observation of the Scale Relation between Density of States and Pairing Gap in a Dirty Superconductor*

Chang-Jiang Zhu[1,2†], Limin Liu[1,2†], Peng-Bo Song[1,2†]. Han-Bin Deng[1,2], Chang-Jiang Yi[1], Ying-Kai Sun[1,2], R. Wu[1,5], Jia-Xin Yin[4], Youguo Shi[1**], Ziqiang Wang[6**], Shuheng H. Pan[1,2,3,5**]

[1]*Beijing National Laboratory for Condensed Matter Physics, Institute of Physics, Chinese Academy of Sciences, Beijing 100190, China*
[2]*School of Physical Sciences, University of Chinese Academy of Sciences, Beijing 100049, China*
[3]*CAS Center for Excellence in Topological Quantum Computation, University of Chinese Academy of Sciences, Beijing 100190, China*
[4]*Laboratory for Topological Quantum Matter and Advanced Spectroscopy (B7), Department of Physics, Princeton University, Princeton, NJ, USA.*
[5]*Songshan Lake Materials Laboratory, Dongguan, Guangdong 523808, China*
[6]*Department of Physics, Boston College, Chestnut Hill, MA, USA.*

**Abstract**

Theories and experiments on "dirty superconductors" are sophisticated but important for both fundamentals and applications. It becomes more challenging when magnetic fields are present, because the field distribution, the electron density of states, and the superconducting pairing potentials are nonuniform. Here we present tunneling microspectroscopic experiments on NbC single crystals and show that NbC is a homogeneous dirty superconductor. When applying magnetic fields to the sample, we observe that the zero-energy local density of states and the pairing energy gap follow an explicit scale relation proposed by de Gennes for homogeneous dirty superconductors in high magnetic fields. Surprisingly, our experimental findings suggest that the validity of the scale relation extends to magnetic field strengths far below the upper critical field and call for new nonperturbative understanding of this fundamental property in dirty superconductors. On the practical side, we use the observed scale relation to drive a simple and straightforward experimental scheme for extracting the superconducting coherence length of a dirty superconductor in magnetic fields.

**Introduction**

In many important applications of superconductors, such as lossless electric power transmission and superconducting magnets for MRI and particle accelerators, superconducting materials are not ideal single crystals. Instead, they often contain impurities including chemical impurities and physical impurities (defects and inhomogeneities), or can even be in the amorphous state or alloys. Chemical impurities or defects have also been purposely introduced into superconductors to control the pinning effects of vortices for better performance in magnetic fields. Therefore, the study of these so-called "dirty" superconductors has its importance in both fundamentals and applications. In his celebrated work "Theory of Dirty Superconductors" in 1959, P. W. Anderson studied the striking experimental fact that BCS superconductivity is often insensitive to enormous amounts of nonmagnetic physical and chemical impurities. He showed that the BCS pairing interaction plays a much more unique role in the scattered states than it does in pure superconductors [1]. Dirty Type-II superconductors in high magnetic fields are much more challenging for theoretical calculations and experimental investigations. One of the complications is that the density of states, the pairing potential, and the magnetic field distribution are nonuniform due to the formation of magnetic vortices. In 1964, P. G. de Gennes studied the "Behaviors of Dirty Superconductivity in High Magnetic Fields" [2] in the diffusive regime where the normal state mean field path $l \ll \xi$ the superconducting coherence length. The hallmark of this theoretical work is a remarkable prediction: When a dirty type-II superconductor is placed in high magnetic fields of its mixed state, the local electron density of states depends only on the pair potential at the same position, regardless of the local magnetic field. Practically, this theoretical prediction can be expressed in a very concise and explicit scale form as

$$N(r)/N_0 = 1 - (\Delta(r)/\Delta_0)^2, \qquad (1)$$

where, $r$ is the distance from the center of the vortex, $N(r)$ is the local density of states at zero-energy at position $r$ and $N_0$ is the normal state density of states, $\Delta(r)$ is the superconducting energy gap and $\Delta_0$ is the energy gap in zero field [2, 3]. It is important to note that de Gennes' derivation of the scale expression (1) was made in the critical regime where the magnetic field strength is close to $H_{c2}$. Therefore, $\Delta(r)/\Delta_0$ can be treated as an expansion parameter and high order corrections to the scale form are expected. This explicit scale expression of conceptual significance to the understanding of dirty superconductors in the mixed state should be amenable to local probes of the density of states and the pairing energy gap. However, to our knowledge, such a remarkable prediction of the scaling behavior has not been directly observed experimentally [4].

*Project supported by the Foundations in the Acknowledgement.
**Correspondence authors. Email: span@iphy.ac.cn; ziqiang.wang@bc.edu; ygshi@iphy.ac.cn
†Authors equally contributed to this work.

The direct experimental demonstration of this prediction requires spectroscopic measurements with spatial resolution. Scanning Tunneling Microspectroscopy (STM/S) is known as a spectroscopic technique with high spatial and energy resolution. It has been successfully employed in the investigation of electronic structures of superconductors, particularly in the exploration of the effects of impurities, magnetic vortices, and microscopic inhomogeneities in unconventional superconductors, in which all these effects are very local [5-13]. In this article, we present a direct experimental study of this theoretical prediction by performing STM/S experiments on the transition metal carbide superconductor NbC that has some excellent properties for applications, such as high hardness, high melting point, corrosion resistance, etc. [14-16]. The superconductivity of NbC was first reported in 1952 and its critical temperature was found to be about 11 K, which is higher than that of the elemental Nb [17-19]. Early planar junction tunneling experiments on the NbC films prepared by laser ablation concluded that NbC is a clean superconductor with an intermediate electron-phonon coupling [20]. The submillimeter electrodynamic measurement on such thin-film also demonstrated that the electrodynamic parameters of NbC agree with that derived from BCS theory [21]. Recently, several theoretical and experimental works on this material found evidence for a topological band structure and possible unconventional superconductivity. These findings generated renewed interest in NbC in quantum material research [22, 23].

Our measurements show that the STM/S tunneling spectrum in the superconducting state is spatially homogeneous across the NbC surface, irrespective of the "dirtiness", such as various disorders and the normal state electronic inhomogeneity visible in the topography. This suggests that the NbC studied is a diffusive superconductor consistent with the class of "homogeneous, dirty superconductors" envisioned by de Gennes. Moreover, we found that the Caroli-de Gennes-Matricon (CdGM) vortex core bound states, usually seen in "clean" superconductors, are absent inside the isotropic and well-ordered Abrikosov vortices in a perpendicular magnetic field. Scanning outward from the vortex core center, we find that the zero-bias local density of states $N(r)$ decreases continuously along with the increase of the local pairing gap $\Delta(r)$, indicating a gradual recovery of superconductivity. We find that such spatial evolution agrees remarkably well with the scale form predicted by de Gennes in Eq. (1). Based on these observations, we derive a simple experimental scheme for extracting the superconducting coherence length $\xi$ in the mixed state of dirty superconductors.

**Experiment**

The single crystals of NbC used in this work were grown from Co flux. Starting materials with the molar ratio of Nb:C:Co = 1:1:9 were put in an alumina crucible and heated to 1500 °C in an argon-filled furnace for 20 hours. Then the furnace was cooled slowly down to 1300 °C at a rate of 1 °C/h and finally cooled down naturally to room temperature. The single crystals were separated by washing off the excrescent Co flux with hydrochloric acid[23]. All STM/S measurements were carried out using a home-built ultrahigh vacuum low-temperature scanning tunneling microscope (STM) with a base temperature of 1.25 K and a superconducting magnet of 9 T. The tungsten STM tips were made by the electrochemical corrosion method and cleaned by a field emission process against a gold target. The single-crystal sample was precooled in the cryostat, then cleaved at ~10 K before insertion into the microscope for measurements. We have used the standard lock-in technique to acquire the differential conductance spectrum with a modulation voltage $V_{mod}$ of root-mean-square voltage 30-100 μV$_{rms}$, and a frequency of 781.1 Hz.

**Result and Discussion**

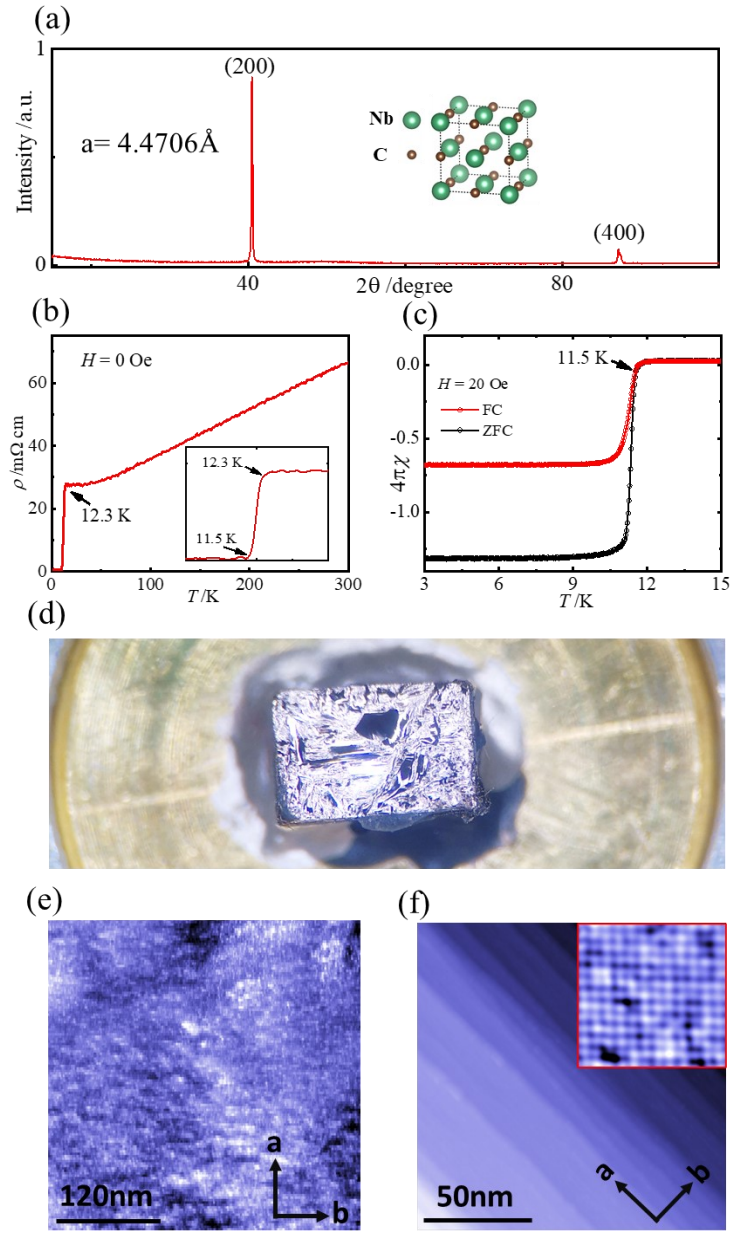

**Figure 1** Structure of single-crystal NbC and the exposed surface after cold-cleaving. **(a)** Result of the XRD measurement with a face-centered cubic structural model of NbC (inset). **(b)** Result of the resistivity measurement. **(c)** Result of the magnetic susceptibility measurement. **(d)** Optical image of the cleaved NbC single crystal sample of 2.5 mm*4 mm in dimension, showing some small mirror-like macroscopic flat areas. **(e)** STM image of a disordered (001) surface ($T = 1.3$ K, $V_s = -100$ mV, $I_t = 10$ pA). **(f)** STM image of a (001) surface with flat terraces ($T = 1.3$ K, $V_s = 100$ mV, $I_t = 1$ nA), the inset is an atomically resolved high-resolution image zoomed in to a flat terrace.

NbC has a face-centered cubic (FCC) crystal structure in the space group of $Fm\bar{3}m$, as shown in Fig. 1(a) (inset). The results of XRD, resistivity and the magnetic susceptibility measurements of the NbC crystals shown in Fig. 1(a), (b), and (c) demonstrate the high quality of the single crystals. The average crystal size chosen for our experiments is a couple of millimeters across. Because of its FCC ionic structure, cleaving these single-crystal samples to achieve flat exposures is rather difficult. By using the *in situ* cold-cleaving technique, we have been able to obtain some macroscopically flat and smooth areas of sufficiently high quality for STM/S measurements. Fig. 1(d) displays an optical image of such a cold-cleaved sample, showing small mirror-like macroscopically flat areas. Most of the time, these flat areas are microscopically disordered, as shown in Fig. 1(e). Occasionally, we observe areas with terraces. In Fig. 1(f), we show an STM image of an area with atomically flat terraces and an atomic-resolution image zoomed in to one of the flat terraces (shown as inset). The diagonal distance of the square lattice in the image is about 4.5 Å, which agrees well with the FCC unit-cell lattice constant of 4.506 Å determined by X-ray diffraction[24]. It is likely that the atomic resolution STM image reveals the carbon lattice, based on the conclusion of the early REED experiment [25].

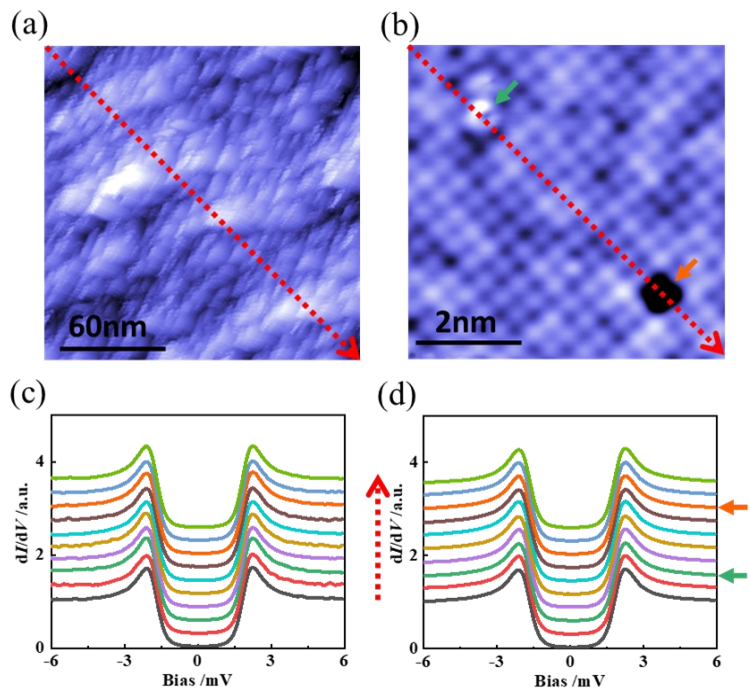

**Figure 2** STM images of two types of surface exposure and the dI/dV spectra showing the almost identical homogeneous superconducting state on the two different NbC surfaces. **(a)** Topographic image of a disordered area. ($T = 1.3$ K, $V_s = 100$ mV, $I_t = 10$ pA). **(b)** Atomically resolved topographic image. The green and orange arrows label the two types of atomic defects, respectively. ($T = 1.3$ K, $V_s = 10$ mV, $I_t = 1$ nA). **(c)** dI/dV spectra taken along the red dotted line marked in (a). ($T = 1.3$ K, $V_s = 10$ mV, $I_t = 1$ nA, $V_{mod} = 100$ mV$_{rms}$). **(d)** dI/dV spectra taken along the red dotted line marked in (b). (Same measuring conditions as for (e)). The spectra are off-set vertically for clarity.

As illustrated above, we are able to expose two types of flat areas, both suitable for STM experiments. One of these has a high degree of roughness and is not possible for atomic-resolution imaging, as shown in Fig. 2(a). The other one is atomically flat and capable for atomic-resolution imaging. Fig. 2(b) is a high-resolution image of such an area. It can be seen that in addition to the resolved atomic lattice, the image shows point defects on the surface as well. The defect labeled by the green arrow appears to be an adatom residing at the interstitial site and the one labeled by the orange arrow an atomic vacancy on the surface. In fact, they could all be chemical impurities. It should also be pointed out that the regions without defects still do not appear to be uniform, indicating the normal density of states (DOS) is electronically inhomogeneous. Fig. 2(b) shows the disordered surface with higher roughness. Judging by the quality of the crystal, these disorders are most likely due to debris, defects, and surface reconstructions resulting from cleaving the ionic crystal. We have carried out scanning tunneling spectroscopic (STS) measurements on both types of surface areas. Fig. 2(c) and Fig. 2(d) plot the $dI/dV$ spectra acquired at a temperature of 1.3 K and along the red dotted lines marked in Fig. 2(a) and 2(b) respectively. All spectra show a U-shaped superconducting energy gap feature with well-defined coherence peaks. Remarkably, the spectra on these two different types of surfaces are almost identical and extremely uniform, demonstrating the superconducting state of NbC is homogeneous and robust against impurity, disorder, and electronic inhomogeneity. These behaviors are in striking contrast to those observed in unconventional superconductors[5-7, 11-13]. On one hand, the fact that all spectra are uniform and identical, even with the ones taken on top of the point defects (impurities), could be used to argue for the crystal sample being free of residual Co atoms (if any, it could only be a very minuscule amount). Otherwise, we would have seen the signatures of Yu-Shiba-Rusinov states in the superconducting tunneling spectra taken on the Co impurities, a phenomenon ubiquitous for magnetic impurities in superconductors [26-29]. On the other hand, the uniformity and robustness of the tunneling spectra could also be regarded as indications of very dirty superconductors with nonmagnetic disorders [1] and a superconducting coherence length much larger than the mean-free path.

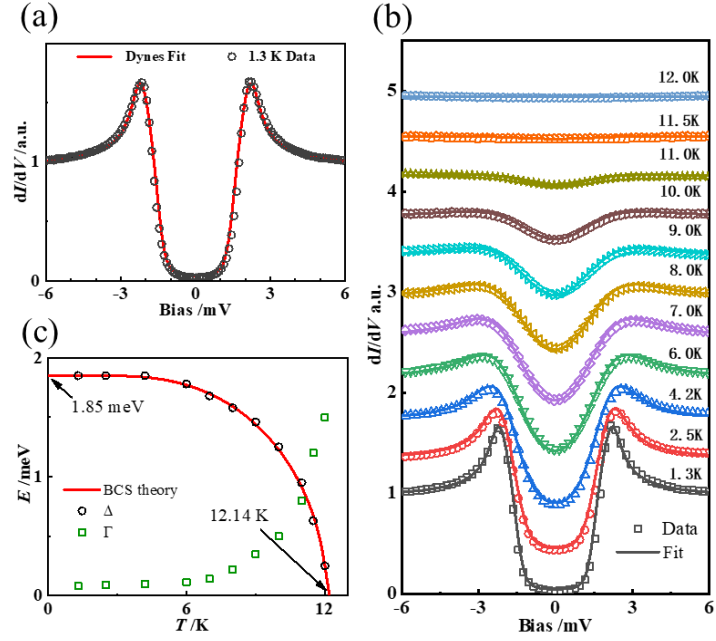

**Figure 3** Temperature dependence of the d*I*/d*V* spectrum. **(a)** Averaged d*I*/d*V* spectrum fitted with Dynes equation ($T = 1.3$ K, $V_s = 10\ mV$, $I_t = 1$ nA, $V_{mod} = 100$ mV$_{rms}$). The circular marks represent the experimental data taken at 1.3 K. Red solid curve represents the theoretical fitting with the Dynes equation. **(b)** d*I*/d*V* spectra as a function of temperature. Solid curves are results of the fit with the Dynes equation. **(c)** Temperature dependence of the superconducting gap energy Δ and the decoherence term Γ. The black solid curve is the fit to the BCS theoretical prediction.

Since the spectra at different locations are almost identical, it is reasonable to fit the normalized average d*I*/d*V* spectrum with a single BCS s-wave gap function using the Dynes equation[30] expressed as

$$N_s(E) = Re\left(\frac{(E-i\Gamma)}{\sqrt{(E-i\Gamma)^2-\Delta^2}}\right). \quad (2)$$

In the equation, $N_s(E)$ is the density of states (DOS) at energy *E*, Δ is the magnitude of the superconducting energy gap, and Γ is the decoherence term, which is inversely proportional to the quasi-particle lifetime and damps the coherence peaks in the tunneling spectrum. In Fig. 3(a), the dark open circles represent the normalized average data of the superconducting spectrum measured at 1.3 K and the red curve is the fitting result. The almost perfect fit of the experimental data gives an isotropic superconducting energy gap magnitude Δ=1.85 meV at 1.3 K and a decoherence term Γ=0.08 meV, slightly larger than the gap value of 1.81 meV obtained from a planner junction measurement[20]. Fig. 3(b) plots the temperature-dependent d*I*/d*V* spectra also with the Dynes fit (solid lines). It can be seen that all the spectra can be fitted to an isotropic BCS gap function extremely well. It can also be observed that as the sample temperature is raised, the energy gap decreases and the coherence-peak features in the spectra are broadened. Fig. 3(c) depicts the temperature dependence of the superconducting energy gap, where the black open circles mark the superconducting gap values and the green open squares mark the values of the decoherence, all extracted from the fit. When temperature increases, the superconducting gap $\Delta(T)$ first decreases slowly, and then drops rapidly, while the decoherence term $\Gamma(T)$ increases monotonically. As also depicted in Fig. 3(c), the temperature dependence of the energy gap follows the BCS theory (solid red line) very closely. There is certainly a valid concern about the error in the fitting process. One can use a normalized $\chi^2$ as a measure of the quality of the fit[31]:

$$\chi^2 = \frac{1}{n-p}\sum_{i=0}^{n}\frac{(x_i-f_i)^2}{x_i}, \qquad (3)$$

where $n$ is the number of data points in the curve, $p$ is a fitting parameter, $x$ is the measured data value, and $f$ is the fit value.

Using this error analysis as the guide for the fit, we obtain the best fits for all the spectra and find that the $\chi^2$ values for all our fits are less than 0.2%, indicating a very small fitting error. We can thus confidently extrapolate the transition temperature $T_c$ to be of 12.14 $\pm$ 0.02 K, which is higher than the bulk $T_c$ of 11.5 K determined by the magnetic susceptibility measurement. Considering STM/S is a surface technique specifically sensitive to the electronic properties of the surface, the higher $T_c$ obtained in our STS measurement can be reasonably ascribed to the surface enhancement of superconductivity. This reasoning is quite plausible, since our resistivity measurement gives an onset $T_c$ of 12.3 K, higher than the bulk $T_c$ of 11.3 K from susceptibility measurement, as shown in Fig. 1(b) and (c). That is because if the surface superconducts at a higher temperature $T_c$, it will shunt the bulk resistance and lead resistivity measurement to give transition temperature as the surface $T_c$. Whereas in the planer junction experiment[20], the surface exposure was covered by insulating barrier, and therefore the measurement resulted the bulk $T_c$ of 11.3 K. To the low-temperature end of the fit, the spectrum of the density of states has much more pronounced quasiparticle features, such as very deep gap with flat bottom and very sharp coherence peaks, hence there will be much less margin for miss fit. Therefore, we can be more confident to extrapolate the zero-temperature gap value $\Delta_0$ =1.85 meV. With these values we can calculate the ratio of the superconducting energy gap $\Delta_0$ and $T_c$, an indicator of the Cooper-pair coupling strength, to be of $2\Delta_0/k_BT_c$ = 3.54 that is very close to the BCS weak coupling value of 3.53.

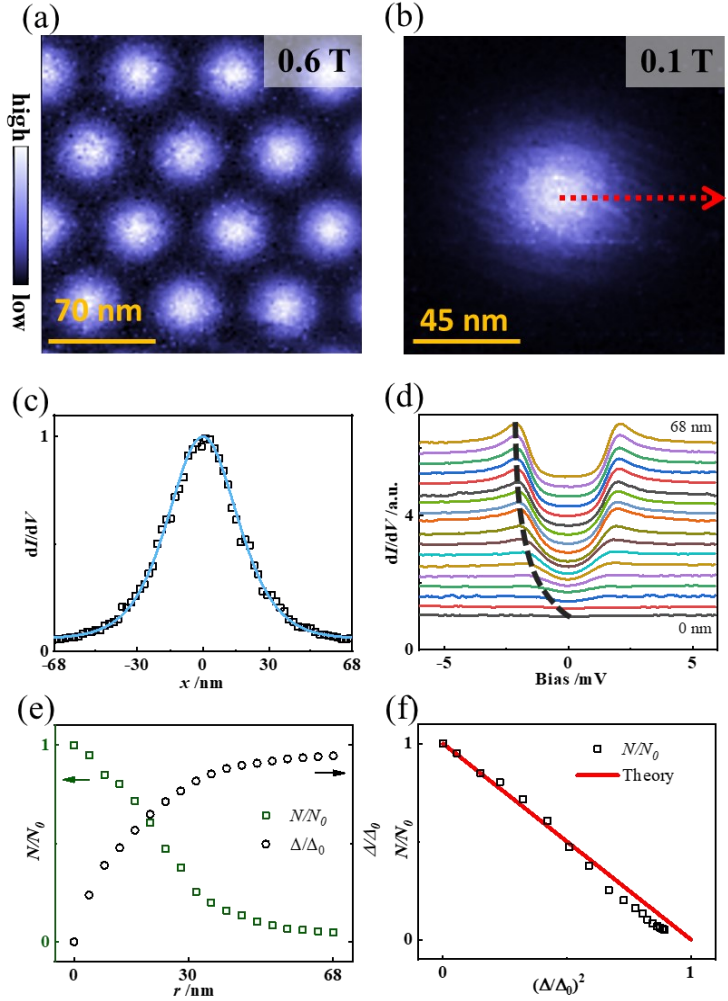

**Figure 4** Magnetic vortex imaged on the surface of NbC. The brightness in the images is proportional to the differential tunneling conductance. **(a)** Image of the Abrikosov vortex lattice when the sample is in a 0.6 T magnetic field perpendicular to the surface. ($T = 1.3$ K, $V_s = 10$ mV, $I_t = 1$ nA, $V_{mod} = 100$ mV$_{rms}$). The solid light blue curve is the fit with the scale expression. **(b)** Zero bias d$I$/d$V$ map in a 0.1 T magnetic field showing a single vortex at the center. **(c)** Line-profile of the zero-bias conductance through the center of the vortex. **(d)** Spatial evolution of the d$I$/d$V$ spectrum along the red dotted line marked in (b). The black dashed curve provides a guide to the eye, showing the evolutions of the energy gap and the coherence peak. **(e)** Spatial dependence of the normalized superconducting energy gap and the normalized zero-energy density of states as a function of the distance from the center of the vortex. **(f)** Scale relation between the normalized zero-energy density of states and the normalized superconducting energy gap.

The microscopic and spectroscopic measurements described above all show very conventional BCS-like behaviors with possible indications of NbC being a dirty superconductor; no signatures of unconventional superconductivity have been detected. Magnetic fields break time-reversal symmetry, so have often been used to detect the signature of unconventional superconductivity, particularly with impurities or magnetic vortices [32, 33]. First, we apply a magnetic field of 0.6 T (half of the critical field value

$H_{c2}$ = 1.21 T[23]) perpendicular to the measured sample surface. No impurity effects have been found. Instead, a well-ordered hexagonal Abrikosov vortex lattice can be clearly observed in the zero-bias d$I$/d$V$ map, as displayed in Fig. 4a, confirming that NbC is a type-II superconductor. The observed vortices all have an isotropic round shape that is consistent with the s-wave pairing symmetry. As we reduce the magnetic field to 0.1 T, the density of the vortices decreases as expected, and the distance between the neighboring vortices increases to about 136 nm. Then, we zoom in to a single vortex to study the electronic structure of the vortex in greater detail. Fig. 4(b) displays a high-resolution image of this single vortex. Actually, this image is a 135 ×135 nm zero-bias conductance (ZBC) map with a single vortex in the center, therefore it readily gives the high-resolution ZBC line profile through the vortex (Fig. 4(c)). In Fig. 4(d) we show the spatial evolution of the d$I$/d$V$ spectra along the red dotted line marked in Fig. 4(b). It can be seen that the tunneling spectrum taken at the center of the vortex shows a featureless normal state spectrum. While stepping away from the vortex center, the superconductivity gradually recovers, indicated by the gradual development of the superconducting energy gap and the coherence peak features. Unexpectedly, we do not observe the Gaussian-shaped piling-up peak of the Caroli de Gennes Matricon (CdGM) vortex core states, which is usually seen in the vortex of a conventional BCS type-II superconductor[34-36]. The absence of the CdGM state peak is commonly imputed to the enormous amount of electron scattering by impurities or defects, particularly when the superconductor is in the "dirty limit" judged by comparing the electron mean-free-path $l$ in the material with the coherence length $\xi$ of the superconducting state [37]. Our NbC single crystals have been characterized to show the electron-mean-free path is about 3.3 nm, much smaller than the coherence length of 16.5 nm [23]. Therefore, our superconducting NbC samples can be considered as in the "dirty limit", and the absence of the CdGM states may be due to impurity scattering in the normal region of the vortex core. As discussed in the experimental results without magnetic fields, the superconducting state is homogeneous and robust against disorders. The spectroscopic results of our experiments show that the low-temperature decoherence term Γ, that partially accounts for the scattering effects, is rather small, only about 4% of the superconducting gap energy. This low decoherence thus indicates the scattering of the quasi-particles is diffusive, although the precise nature of the diffusive scattering is unclear at present. The decoherence term becomes comparable and greater than the gap energy only as the temperature approaches the superconducting transition temperature (shown in Fig. 3(c)), while our measurements (with or without the magnetic fields) are performed at 1.3 K, a temperature far below the critical temperature.

Based on the experiments and results detailed above, we believe that our NbC crystals are the type of homogeneous, dirty superconductors with diffusive electrons considered in de Gennes' theoretical work. The absence of the CdGM vortex core states in our spectroscopic results further encourages a direct comparison to de Gennes' theory that does not account for the CdGM states. From the spectra shown in Fig.4(d), we can directly take the values of the zero-energy local density of states *N(r)* and easily extract the values of the local superconducting gap energy by the fitting procedure using Dynes formula described earlier. The normalized results are shown in Fig.4(e). We then re-plot these normalized results in Fig.4(f) with the coordinates matching the expression of a linear scale form. Obviously, the experimental results agree well with the theoretical prediction in Eq. (1). The agreement is in fact rather remarkable since the applied field strength in our experiments is only 0.1 T, far less than the upper critical field of $H_{c2}$ =1.21 T of NbC [23], whereas the theoretical scale expression is derived under the condition of subcritical fields $(H_{c2} -H)/H_{c2} \ll 1$ as well as the truncation of all higher-order terms in $\Delta(r)/\Delta_0$. One may argue that the unexpectedly small deviation is caused by corrections due to the low applied magnetic field in the experiments compared to the theoretical result obtained near $H_{c2}$ and the higher-order terms in the perturbative expansion. Alternatively, it may also be due to the overlap of neighboring vortices even at such a small field, which causes the local density of states to become higher and the local energy gap smaller. Further experiments with even lower fields (above the very small $H_{c1}$ of NbC ) are desirable to study the change of the deviation systematically.

Using the scale property of the density of states to the superconducting energy gap, we can derive a simple experimental scheme to acquire the coherence length $\xi$ in magnetic fields. Usually, to estimate $\xi$, the normalized line profile of the density states $N(r)$ is fitted empirically to the exponential function $(1/r)\cdot\exp(-r/\xi)$[38] or to the function $1-\tanh(r/\xi)$[39] that includes the Ginzburg-Landau expression of the pairing order parameter[40]. However, neither procedure fits the experimental data well, not even near the center of the vortex where the field is close to the upper critical field $H_{c2}$. We show now that the demonstrated de Gennes' scale expression (1) can be used for a more reliable estimate of $\xi$. Using the Ginzburg-Landau expression for the normalized order parameter (energy gap) $\Delta(r) \sim \tanh(r/\xi)$, we obtain the formula

$$N(r)/N_0 = 1 - b \cdot (tanh\ (r/\xi)/\Delta_0)^2. \tag{4}$$

In this formula, b is a fitting parameter to account for the vortex-overlapping effect that causes the measured gap value to be smaller and the density of states value to be larger than the true ones of a single vortex far away from the others in the vortex lattice. Using the formula (4) to fit the data in Fig. 4(c), we find that the fit is nearly perfect (blue curve in Fig. 4(c)) and gives $b = 0.942 \pm 0.002$ and $\xi = 21.3 \pm 0.13$ nm for the external field of 0.1 T. Factor b scales the energy gap value by $\sqrt{b} = 0.971$. The normalized gap value measured at position $r = 68$ nm (middle position between the two nearest vortices) is 0.945 as shown in Fig. 4(e). It is fairly close to the expected value of 0.971, demonstrating the vortex-overlapping effect on the measurement. If we further consider the vortex-overlapping effect on the density of states at the center of the vortex is almost negligible, and the measured density of states at position $r =68$ nm is the sum of the contributions from both nearest neighboring vortices, we can thus take half of the density of states value and force-fit the scaling equation (3) through only three data points (-68 nm, $N(-68\ nm)/2N_0$), (0, 1) and (68 nm, $N(68\ nm)/2N_0$). Then we obtain the coherence length $\xi = 15.5 \pm 2.9$ nm, quite close to the value of 16.3 nm measured using much more sophisticated instruments and analytical schemes [23].

**Conclusion**

To summarize, we have performed a high-resolution STM/S investigation of the superconducting state on the surface of NbC single crystals. We have found that, despite the strong disorder implying a disordered normal state, NbC is a spatially homogeneous type-II s-wave superconductor at low temperatures in the class of "dirty" or diffusive superconductors. We demonstrate that the high-resolution STS allows a direct study of the concurrent spatial evolution of the local density of states and the paring energy gap as a function of the distance to the center of the vortex core. We find that the two quantities are closely related by the scale form proposed by de Gennes for dirty type-II superconductors with diffusive electrons, providing the first direct experimental demonstration of this theoretical prediction. Surprisingly, our experimental findings suggest that the validity of the scale form extends far beyond the regime of perturbative expansion in the critical region near $H_{c2}$ treated in the de Gennes' theory, therefore call for a new theoretical understanding of the fundamental scale behavior of dirty superconductors.

This work is supported by the National Natural Science Foundation of China (Grant No. 11227903), the Beijing Municipal Science and Technology Commission (Grant Nos. Z181100004218007 and Z191100007219011), the National Basic Research Program of China  (Grants No. 2015CB921304), National Key Research and Development Program of China (Grant Nos. 2017YFA0302903, 2016YFJC010282, and 2016YFA0300602) and the Strategic Priority Research Program of Chinese Academy of Sciences ( Grant Nos. XDB07000000, XDB28000000, and XDB33000000). National Natural Science Foundation of China (Grants No. 12004416, and No. U2032204), the Ministry of Science and Technology of China (Grants No. 2016YFA0300604). Z.W. is supported by the U.S. Department of Energy, Basic Energy Sciences Grant No. DE-FG02-99ER45747.